\documentclass[makeidx]{llncs} 

\usepackage{pst-node,amssymb,url,enumerate,cite}

\usepackage[british]{babel}
\usepackage[utf8]{inputenc}

\newcommand{\bra}[1]{\ensuremath{\left\langle#1\right|}}
\newcommand{\ket}[1]{\ensuremath{\left|#1\right\rangle}}

\title{Halting in quantum Turing computation}

\author{Willem Fouch\'e\inst{1} \and Johannes Heidema\inst{2} \and Glyn Jones\inst{3} \and Petrus H. Potgieter\inst{4}}
\institute{Department of Decision Sciences, \email{fouchwl@unisa.ac.za}
\and Department of Mathematical Sciences, \email{heidej@unisa.ac.za}
\and Department of Physics, \email{joneseg@unisa.ac.za}
\and Department of Decision Sciences, \email{php@member.ams.org}\\
\medskip
University of South Africa, PO Box 392, Unisa 0003, Pretoria
}

\makeindex

\begin{document}

\maketitle

\begin{abstract}
The paper considers the {halting scheme for {quantum Turing machine}s}. The scheme originally proposed by {Deutsch} appears to be correct, but not exactly as originally intended. We discuss the result of Ozawa \cite{Oza97a} as well as the objections raised by Myers \cite{My97}, Kieu and Danos \cite{kieu_no-go_2001} and others. Finally, the relationship of the halting scheme to the quest for a {universal {quantum Turing machine}} is considered.
\end{abstract}

\index{quantum Turing machine}
\index{quantum Turing machine!halting scheme for}
\index{quantum computing}
\index{quantum Turing machine!universal}

\section{Introduction}

The quantum logic circuit model, initiated by Feynman and {Deutsch}, has been more prominent than {quantum Turing machines} in research into quantum computing. However {quantum Turing machine}s have been studied for at least two major reasons:
\begin{enumerate}[(a)]
\item Quantum Turing machines form a class closely related to deterministic and probabilistic Turing machines, the basis for the theory of classical computation which is well developed and quite well understood.
\item Quantum logic circuits are devices purpose built for one specific application whereas {quantum Turing machine}s are devices that (like ordinary Turing machines) can be used to operate on arbitrarily large inputs and that can (possibly---this depends on the existence of a universal device in the same class) be programmed.  Programmability also allows a program-based notion of complexity.
\end{enumerate}
Since a {quantum Turing machine} (QTM) could in principle operate for an indefinite number of steps, the interaction between the operator/observer and the apparatus is of crucial importance. Specifically, the operator needs to be able to tell when s/he may disturb the device and observe the output. This makes the halting of a {quantum Turing machine} a quite delicate issue (as also mentioned in \cite{Hirvensalo02} for example) and this paper examines the solution proposed by {Deutsch}, which is part of the standard description of the QTM.

\section{Classical and probabilistic Turing machines}

Since {quantum Turing machine}s are based on the ordinary Turing machine, we start by reviewing the classical model. 
By the beginning of the twentieth century mathematicians had become quite interested in establishing a formal model of computability. In response,  Alan Turing described an abstract device in 1936, now called a \textit{Turing machine}, which follows a simple, finite set of rules in a predictable fashion to transform finite strings (input) into finite strings (output, where defined). The Turing machine (TM) can be imagined to be a small device running on a two-way infinite tape with discrete cells, each cell containing only the symbol \textbf{0} or \textbf{1} or a blank. The TM has a finite set of possible internal states and a head that can read the contents of the cell of the tape immediately under it. The head may also, at each step, write a symbol to the cell over which it finds itself. There are two special internal states: an \textit{initial state} $q_0$ and a \textit{halt state} $q_H$.

A TM has a finite list of instructions, or \textit{transition rules}, describing its operation. There is at most one transition rule for each combination of cell content (under the head) and internal state. If the internal state is $q_{i}$ and the head is over a cell with content $S_{j}$ then the machine looks for a rule corresponding to $(q_i, S_j)$. If no rule is found, the machine enters the halting state immediately. If a rule corresponding to $(q_i, S_j)$ is found, it will tell the machine what to write to the cell under the head, whether to move left or right and which internal state to enter. There is no transition rule corresponding to the halting state. Sometimes we refer to the entire collection of individual rules for all the different $(q_i, S_j)$ as \textit{the transition rule} of the machine. 
A \textit{computation} consists of starting the TM with the head over the first non-blank cell (which we may label position 0 on the tape) from the left of the tape (it is assumed that there is nothing but some finite \textit{input} on the tape) and the machine in internal state $q_0$. Now the transition rules are simply applied until the machine enters the halting state $q_H$, at which point the content of the tape will be the \textit{output} of the computation. If, for some input, the machine never halts then the output corresponding to that input is simply undefined. It is clear how every TM defines a (possibly, partial) function $f:\mathbb{N}_0\rightarrow\mathbb{N}_0$  from the set of counting numbers to itself.

Turing machines are the canonical models of computing devices. No deterministic device, operating by finite (but possibly unbounded) means has been shown to be able to compute functions not computable by a Turing machine. In fact, one may view one's desktop computer as a Turing machine with a \textit{finite} tape. 

A probabilistic Turing machine (PTM) is identical to an ordinary Turing machine except for the fact that at each machine configuration $\left(q_i,S_j\right)$ there is a finite set of transition rules (each with an associated probability) that apply and that a random choice determines which rule to apply. We fix some threshold probability greater than even odds (say, 75\%) and say that a specific PTM computes $f(x)$ on input $x$ if and only if it halts with $f(x)$ as output with probability greater than  75\%. 

\section{Operation of a {quantum Turing machine} (QTM)}
\label{sec:def}

The \textit{quantum Turing machine} (QTM) was first\footnote{Paul Benioff had related a similar idea somewhat earlier \cite{Ben80a} but primarily in connection with presenting a possible physical basis for reversible computing.} described by David {Deutsch} \cite{Deu85a}.  The basic idea is quite simple, a QTM being roughly a probabilistic Turing machine (PTM) with complex transition amplitudes (the squared moduli of which add up to one at each application) instead of real probabilities. 
The QTM is related to the classical deterministic TM in much the same way as the PTM is. 

In the following the \textit{classical machine} is a machine with a two-way infinite tape, starting over position 0 on the tape as described above, that we use as a kind of template for the {quantum Turing machine}. The  corresponding {quantum Turing machine} (QTM) might work as follows (based on the {Deutsch} description \cite{Deu85a}, Ozawa \cite{Oza97a}, Bernstein and Vazirani \cite{BV97a}).
\begin{enumerate}[I.]
\item The quantum state space of the machine is spanned by a basis consisting of states 
$$\ket{h}\ket{q_C}\ket{T_C}\ket{x_C}$$ 
where $\ket{h}$ is the halt qubit, $h\in\{0,1\}$ and $(q_C,T_C,x_C)$ is a configuration of the corresponding classical machine, where $x_C$ denotes the position of the head, $q_C$ the internal state of the machine and $T_C$ the non-blank content of the tape. 
\item Special initial and terminal internal states have been identified (corresponding to the initial state and halting state of the classical machine).
\item The single  transition rule is now a unitary operator $U$ which, in each step, maps each basic $\ket{h}\ket{q}\ket{T}\ket{x}$ to a superposition of only finitely many  $\ket{h'}\ket{q'}\ket{T'}\ket{x'}$, where 
\begin{enumerate}
\item the rule is identical for $\ket{h}\ket{q}\ket{T_1}\ket{x}$ and $\ket{h}\ket{q}\ket{T_2}\ket{y}$ when $T_1$ in position $x$ and $T_2$  in position $y$ have the same content, i.e. the rule depends only on the content of the tape under the head and the internal state $q$ and not on the position of the head or on the content of the rest of the tape;
\item $T'$ and $T$ differ at most in position $x$;
\item $|x'-x|\leq 1$ (depending on whether the corresponding classical machine moves one position to the left, to the right, or not at all);
\item $h'=1$ if and only if $q'$ is the halting state of the classical machine; and
\item $T'=T$, $q'=q$ and $h'=h$ whenever $h=1$.
\end{enumerate} 
Finitely many subrules
\begin{equation}
\label{eqtr}
\ket{h}\ket{q}\ket{T}\ket{x} \quad \longmapsto \quad \sum_{i=1}^n c_i \ket{h_i}\ket{q_i}\ket{T_i}\ket{x_i}
\end{equation}
will determine $U$ as there are, by the stipulations above, only finitely many possible---given that the alphabet of the tape (binary in our case) and the number of internal states are both finite.
Note that the transitional rule (``program'') will have a finite specification only if the transition amplitudes in the superposition of the $\ket{h'}\ket{q'}\ket{T'}\ket{x'}$ are all \textit{computable} complex numbers, which we will of course assume to be the case throughout. The transition rule can also, obviously, be extended (linearly) to finite superpositions of $\ket{h}\ket{q}\ket{T}\ket{x}$.
\item The machine is started with a finite superposition of inputs
$$\ket{0}\ket{q_0}\ket{T}\ket{x}.$$
Because of the form that the transition rule is  allowed to take (and the fact that there are only finitely many internal machine states) the machine will be in the superposition of  only \textit{finitely many} basic states $\ket{h}\ket{q}\ket{T}\ket{x}$ at any step during the entire run\footnote{A more hazy concept than for classical Turing machines, as a QTM only really stops when one has observed the halt qubit and the content of the tape, so one may think of the transition rule being applied \textit{ad infinitum}, step-by-step, unless the operator (physically, classically and externally) stops the machine.} of  computation.
\end{enumerate} 
The description of the machine given here differs from a classical reversible Turing machine in two obvious respects.
\begin{enumerate}
\item Transition rules are allowed to map a state of the machine to the superposition of several states. The crucial distinction with classical probabilistic machines is that the QTM goes to a quantum superposition of states whereas the classical PTM can be seen as either going to a classical probability distribution over states or to a specific state with some classical probability. Quantum computing, of course, uses superposition in an essential way\footnote{An very readable and accessible explanation of how and why this works can be found in \cite{fortnow_one_2000}.}---as in the the algorithms of Shor or Grover.
\item The input is allowed to be a superposition of a finite number of ``classical'' inputs.
\end{enumerate}
It is not immediately obvious why a finite collection of specifications of the form (\ref{eqtr}) should necessarily define a unitary $U$, however, just as it might not be apparent why a finite collection of rules 
$$\ket{h}\ket{q}\ket{T}\ket{x} \quad \longmapsto \quad \ket{h'}\ket{q'}\ket{T'}\ket{x'}$$
for a classical machine would necessarily specify a reversible machine. Unitarity is, of course, a precondition for the quantum device to be feasible.

\section{Physicality of the QTM}
\label{sec4}

We start by assuming that the operator $U$ described above is what is often called \textit{well-formed}, i.e. that the subrules of the form (\ref{eqtr}) give rise to a unitary operator. 
Without loss of generality everything can be assumed to be coded in binary so that each position on the tape will 
correspond to a single qubit (quantum bit).  A unit of quantum information, the qubit is a two level quantum mechanical system, whose state is described by a linear superposition of two basis quantum states, often labelled \ket{0} and \ket{1}.
The actual (quantum) state space of the machine will be a direct sum of $n$-qubit spaces (where $n$ is an indication of how much tape has been used, each $n$-qubit space being the $n$-fold tensor of the single qubit space). 
The direct sum is, however, not a \textit{complete} inner-product space (i.e. not a Hilbert space) and therefore---by the postulates of quantum mechanics---not a valid state space. However, the underlying Hilbert space can be taken to be the completion of the direct sum and a unitary operator $U$ on the direct sum (see \cite{BV97a}) can be extended to a unitary operator $\widehat{U}$ on the Hilbert space. This completed space and operator will correspond to the physical system associated with the QTM, thereby taking care of the \textit{physicality} of the QTM, provided that $U$ was \textit{well-formed}.

To see that well-formedness of $U$ is not a triviality, consider a putative QTM with initial state $q_0$ and final state $q_H$. Let the following subrules define the machine:
\begin{eqnarray*}
\ket{0}\ket{q_0}\ket{\ldots 0\ldots}\ket{x} & \quad \longmapsto \quad &\frac{1}{\sqrt{2}} \ket{1}\ket{q_H}\left(\ket{\ldots 0\ldots}+\ket{\ldots 1\ldots} \right)\ket{x+1} \\ \\
\ket{0}\ket{q_0}\ket{\ldots 1\ldots}\ket{x} & \quad \longmapsto \quad & \ket{1}\ket{q_H}\ket{\ldots 1\ldots}\ket{x+1} \\ \\
\ket{1}\ket{q_H}\ket{T}\ket{x} & \quad \longmapsto \quad & \ket{1}\ket{q_H}\ket{T}\ket{x+1}
\end{eqnarray*}
These subrules specify that the machine starts out by replacing a 0 under the head by a superposition of 0 and 1, leaves a 1 initially under the head unchanged and then halts (i.e. enters a final state in which the tape head only may still move). Given a classical input, it could be said to operate reversibly as we can recover the input from the output. However, the putative QTM is not \textit{well-formed} and therefore not a real QTM at all since 
$$\ket{0}\ket{q_0}\ket{\ldots 0\ldots}\ket{x} \quad \mbox{is perpendicular to} \quad \ket{0}\ket{q_0}\ket{\ldots 1\ldots}\ket{x}$$
but
$$ \frac{1}{\sqrt{2}} \ket{1}\ket{q_H}\left(\ket{\ldots 0\ldots}+\ket{\ldots 1\ldots} \right)\ket{x+1}$$
is certainly not perpendicular to $\ket{1}\ket{q_H}\ket{\ldots 1\ldots}\ket{x+1}$ at all. A possibly correct version of the same machine could perhaps be given by
\begin{eqnarray*}
\ket{0}\ket{q_0}\ket{\ldots 0\ldots}\ket{x} & \quad \longmapsto \quad &\frac{1}{\sqrt{2}} \ket{1}\ket{q_H}\left(\ket{\ldots 0\ldots}+\ket{\ldots 1\ldots} \right)\ket{x+1} \\ \\
\ket{0}\ket{q_0}\ket{\ldots 1\ldots}\ket{x} & \quad \longmapsto \quad & \frac{1}{\sqrt{2}} \ket{1}\ket{q_H}\left(\ket{\ldots 0\ldots}-\ket{\ldots 1\ldots} \right)\ket{x+1} \\ \\
\ket{1}\ket{q_H}\ket{T}\ket{x} & \quad \longmapsto \quad & \ket{1}\ket{q_H}\ket{T}\ket{x+1}
\end{eqnarray*}
for example, as the previous objection has been eliminated. The constraint of unitarity clearly excludes certain combinations of transition rules and directly implies that a great deal of care should be taken to ensure well-formedness of any machine that one may construct.

\section{Time evolution of the QTM}

If $U$ is the operator that describes one application of the transition rule (i.e. one step in the operation) of the machine, then the evolution of an  unobserved machine (where not even the halt bit is measured) for $n$ steps is simply described by $V=U^n$. If the first measurement occurs after $n_1$ steps, and the measurement is described by an operator $J_1$ then the evolution of the machine for the first $n_1+j$ steps is described by $$U^jJ_1U^{n_1}$$ which is in general no longer unitary since the operator $J_1$ is a measurement (always in the computational basis). It is important to note that the machine evolves unitarily only when no measurement takes place at all. The {quantum Turing machine} should therefore not be seen as a pure quantum device but as a kind of hybrid device. Actually, as will be discussed below, the output from operating the machine for $n$ steps without measurement of the halt bit is equivalent to operating it for $n$ steps and observing the halt bit after every step. However, the QTM is envisioned with no explicit limit on the number of steps for which it may run, so the whole machine is evidently not equivalent to any single quantum experiment.

\section{The halting scheme}

The output of the machine on the tape is of course a superposition of basis states and should be read off after having measured the content of the halt bit and finding it in the state 1. The operator may at any time, or indeed between every two applications of $U$, measure the halt bit\footnote{
The halt \textit{qubit}, of course, until we measure it.} in order to decide whether to read the tape content (and collapse the state of the machine to one of the basis states). The halt bit is intended to give the operator of the machine an indication of when an output may be read off from the tape (and by observation collapsing the system to an eigenstate) without interfering excessively with the computation. 

It seems that {Deutsch}'s original idea was that there would be no entanglement at all between the halt bit and the rest of the machine, but this cannot be guaranteed, as desribed by Myers \cite{My97}. 
The \textit{output} of a QTM for some specific input $x$ (which may be a superposition of classical inputs) is a probability distribution $P_x$ over all possible contents of the tape at the time of observing the halt bit to have been activated. 
Actually, Miyadera and Ohya \cite{miyadera_halting_2003} have provided a simple proof that it is not possible to effectively distinguish between those {quantum Turing machine}s which have deterministic and those which have probabilistic halting behaviour.

\subsection{Validity of the halting scheme}

In an ordinary QTM the evolution of the machine continues even when the halt bit has been observed, 
and can continue after the halt bit has been observed without perturbing the probability distribution that has been defined to be the QTM's \textit{output} (Ozawa \cite{Oza97a}) since the observation projects one, in a certain sense, only into a specific ($h=0$ or $h=1$) branch of the computation. Let us consider this idea in slightly more detail.

Suppose $U$ describes a QTM with a proper halting scheme, as above. By this we mean that $U$ is unitary and that if 
$$\ket{1}\ket{q_H}\ket{T}\sum_i c_i\ket{x_i}$$ 
is a halting configuration that occurs (maybe in a superposition) during the evolution of the machine for some valid input, then 
$$U\ket{1}\ket{q_H}\ket{T}\sum_i c_i\ket{x_i} =\ket{1}\ket{q_H}\ket{T}\sum_j d_j\ket{y_j}$$ 
for some $y$. This simply means that $U$ satisfies condition III(e) of Section \ref{sec:def} which is sufficient to guarantee this. Assume that the QTM is in a state
$$c_1\ket{1}\ket{\phi_1}+c_2\ket{0}\ket{\phi_2}$$
after $n>0$ applications of $U$ and (Scenario 1) that the halt bit is not measured at this point. One may assume $\ket{\phi_1}$ and $\ket{\phi_2}$ to be normalised, of course, and since $\ket{1}\ket{\phi_1}$ and $\ket{0}\ket{\phi_2}$ are orthogonal\footnote{Obviously, the probability of (Scenario 2) measuring an activated halt bit at this stage would be $|c_1|^2$.}, 
$$|c_1|^2+|c_2|^2=1.$$
Apply $U$ once more, to get
$$c_1U\ket{1}\ket{\phi_1}+c_2U\ket{0}\ket{\phi_2}$$
Since $U$ is unitary and because of III(e)
$$U\ket{1}\ket{\phi_1} = \ket{1}\ket{\psi_1}$$
and furthermore
$$U\ket{0}\ket{\phi_2} = d_1\ket{1}\ket{\psi_2}+d_2\ket{0}\ket{\psi_3}$$
for some normalised $\psi_i$ and $d_i$ with 
$$|d_1|^2+|d_2|^2=1.$$
Therefore the state of the device is now 
$$c_1\ket{1}\ket{\psi_1} + c_2d_1\ket{1}\ket{\psi_2}+c_2d_2\ket{0}\ket{\psi_3}.$$
It follows from unitarity of $U$ that since $\ket{1}\ket{\phi_1}$ is orthogonal to $\ket{0}\ket{\phi_2}$ it must also be the case that $U\ket{1}\ket{\phi_1}$ is orthogonal to $U\ket{0}\ket{\phi_2}$. In other words
$$\ket{1}\ket{\psi_1} \quad\mbox{is orthogonal to}\quad d_1\ket{1}\ket{\psi_2}+d_2\ket{0}\ket{\psi_3}.$$
But $\ket{1}\ket{\psi_1}$ is orthogonal to $\ket{0}\ket{\psi_3}$ and hence the three normalised states
$$\ket{1}\ket{\psi_1},\quad \ket{1}\ket{\psi_2},\quad \ket{0}\ket{\psi_3}$$
are pairwise orthogonal. The probability of now measuring an activated halt bit is therefore
$$|c_1|^2+|c_2d_1|^2$$
which is greater than at the previous step, as one might expect. Unsurprisingly, if (Scenario 2) we had measured the halt bit one step earlier then we would have read off an activated state with probability $|c_1|^2$ and an inactive halt bit with probability $|c_2|^2$. Having measured an inactive halt bit would have collapsed the state of the machine to $\ket{0}\ket{\phi_2}$ which, after another appplication of $U$ would give
$$d_1\ket{1}\ket{\psi_2}+d_2\ket{0}\ket{\psi_3}.$$ 
Observing the halt bit now would show an activated state with probability $|d_1|^2$. Classical probability calculus, given that the events are mutually exclusive and using Bayes' formula, yields a probability of having observed an activated halt bit {after} the last step of Scenario 2 as
$$|c_1|^2+|c_2| ^2 \cdot | d_1|^2$$
which is then the same as the probability of measuring it {after} the last step of Scenario 1.

The preceding elementary exposition illustrates the idea developed by Ozawa in \cite{Oza97a}. That paper considers the probability of observing the tape of the machine in a specific state $T$ with the halt bit activated, either (i) by observing the halt bit after $n_1$ steps (and possibly reading of the content of the tape then, if the halt bit was activated) and then again after another $n_2$ steps or by (ii) just letting the machine run without observation and after $n_1+n_2$ steps measuring the halt bit (and possibly the tape). Ozawa showed that the probability of observing the specific state $T$ of the tape is identical in the two cases. It should perhaps be noted that in Ozawa's description the position of the head on the tape is considered part of the tape, whereas it has been explicitly separated in this contribution. This does not detract in any essential way from Ozawa's result.

\subsection{Objections to the halting scheme}

As mentioned earlier, Myers \cite{My97} pointed out that it must be possible to find a QTM in a superposition of basic states with the halt bit activated and basic states with the halt bit not (yet) active. Among other reasons, this is simply true because if $x$ is an input for which the machine takes $N_x$ steps to active the halt bit and $y$ is an input for which it takes $N_y\neq N_x$ steps, then the machine, when the input is a superposition of $x$ and $y$, will---between the $N_x$-th and $N_y$-th step---find itself in such a superposition. This emphasises the inherent probabilistic nature of {quantum Turing machine}s, as also pointed out clearly by Miyadera and Ohya  \cite{miyadera_halting_2003} and is not an objection \textit{per se}, as Ozawa \cite{Oza97a} has shown.

Kieu and Danos \cite{kieu_no-go_2001} purport to have shown the impossibility of any unitary operator $U$ having the desired halting scheme. They argue as follows. Suppose there exists a state $\ket{\psi}$ of the machine such that 
$$\left(\bra{0}\otimes I\right)~ \ket{\psi} =1 $$
where $I$ is the identity operator, and there exists an $N>0$ such that
\begin{equation}\label{eq1}
\left(\bra{1}\otimes I\right) ~ U^N\ket{\psi} >0.
\end{equation}
Let $N$ further be the smallest number for which relation (\ref{eq1}) holds. 
Such a $\ket{\psi}$ could constitute the QTM prepared with a valid input that eventually activates the halt bit---at least a little bit---and $N$ would be the number of applications of $U$ necessary for any chance of observing an activated halt bit. If no such $\ket{\psi}$ exists then the machine described by $U$ has no input leading to a positive probability of an output and is, in other words, useless. Kieu and Danos proceed to observe correctly that (in our notation)
$$U^{N-1}\ket{\psi} \quad \mbox{is perpendicular to all} \quad \ket{1}\ket{q_H}\ket{T}\ket{x}$$
because of the minimality of $N$ and therefore, since $U$ is unitary,
$$U^{N}\ket{\psi} \quad \mbox{is perpendicular to all} \quad U\ket{1}\ket{q_H}\ket{T}\ket{x}.$$
This is then supposed to contradict (\ref{eq1}). It would be true if, as stated in the Kieu and Danos paper,  the subspace $V$ spanned by the  $\ket{1}\ket{q_H}\ket{T}\ket{x}$ were identical to the subspace $\tilde{V}$ spanned by the $U\ket{1}\ket{q_H}\ket{T}\ket{x}$. Kieu and Danos appear to presume that the restriction $U|_V$ of $U$ to $V$ is unitary, from which it would indeed follow that $V=\tilde{V}$, but this presumption is without proof. Of course, $U|_V$ does preserve the inner product but it is not necessarily unitary because its range need not span all of $V$. 
In fact, $U|_V$ being unitary (i.e. $V=\tilde{V}$) would \emph{directly} imply that no $\ket{\psi}$ as above exists. In \cite{kieu_halting_1998} the same assumption (that $U|_V$, denoted $B$ there, is unitary) appears to have lead to the same conclusion. 

Concerns about the halting scheme were raised also by Shi \cite{Shi2002}, albeit in the context originally introduced by {Deutsch},  where superpositions 
$$c_1\ket{1}\ket{\phi_1}+c_2\ket{0}\ket{\phi_2}$$
with $|c_1c_2|>0$ are assumed not to occur. Shi had actually been examining the existence of universal quantum computers where the halting scheme is very relevant to the discourse.

\subsection{The halting scheme and universality}

Consider a general countable class of machines that compute partial functions, i.e. functions that are not necessarily defined for all inputs. Assume that each machine is fully described by a natural number. 
Let $\Phi_n$ denote the random variable (not simply function, since the machine is not assumed to be deterministic) computed by machine $n$ and fix a reasonable 
$$h : \mathbb{N}_0 \times \mathbb{N}_0 \rightarrow \mathbb{N}_0$$
which will be used for the encoding of programs and data as a single input. 
\begin{definition}
\label{deftwo}
 If there exists a number $N$ such that
 \[ \Phi_N \left( h ( n, m ) \right) = \Phi_n ( m ) \]
 which means that the functions are either equal and both defined or both
 undefined if deterministic, and if not deterministic then the values have the same distribution, for all $n$ and $m$, then the machine described by $N$ is called a \textit{universal} machine for the class.
\end{definition}
Programmability follows from universality and this is why universality is such an important concept. 
A universal Turing machine, for example, can simulate all the Turing machines, and is thus  programmable for the entire class of Turing 
machines. In spite of very powerful results by Bernstein and Vazirani \cite{bernstein_quantum_1997}, universality in the sense of the definition above has not (yet) been established \cite{fouche_deutschs_2007}. The observation of the halt bit goes quite far in explaining why not. 
Bernstein and Vazirani showed that there exists a {quantum Turing machine} $\mathcal U$  such that 
\begin{quote}
``for any well-formed\footnote{Meaning that the time evolution operator is unitary, as discussed in Section \ref{sec4}.} 
QTM $M$, any $\varepsilon>0$, and any 
$T$, $\mathcal U$ can simulate $M$ with accuracy $\varepsilon$ for $T$ steps with slowdown polynomial in $T$ and $\frac{1}{\varepsilon}$.''
\end{quote}
The full Bernstein-Vazirani result could be summarised by the statement that
\begin{quote}
there exists a QTM $\mathcal U$ such that  for each QTM $M$ with finite description $\bar{M}$, $n$, $\varepsilon$ and $T$ there is a program ${\mathcal P}(\bar{M},n,\varepsilon,T)$ and a function $f_{\bar{M}}(T,n,\frac{1}{\varepsilon})$  (both recursive in their inputs) such that running $\mathcal U$ on input $\ket{{\mathcal P}(\bar{M},n,\varepsilon,T)}\otimes\ket{x}$ where $|x|=n$ for $f_{\bar{M}}(T,n,\frac{1}{\varepsilon})$ steps results---within accuracy $\varepsilon$---in the same distribution over observable states as running $M$ on input $\ket{x}$ for $T$ steps. \cite{fortnow_one_2000}.
\end{quote} 
So far there is no problem except if one wants to simulate the running of a given QTM using Berstein and Vazirani's $\mathcal U$ for an indefinite period of time, i.e. as long as it might take to obtain a result. In this case $\mathcal U$ would have to be run for $n=100$, reset and run for $n=200$ and so on, for an unknown but finite number of times. The problem is that after each run of $\mathcal U$ (for example, for $n=500$) the resetting to the original input value is no longer necessarily possible---because the observation of the halt bit could actually have caused the familiar measurement-related collapse of the state of the machine\footnote{If not, recovering the initial state (i.e. the input) is as simple as applying the inverse operation $n$ times.} from 
$$c_1\ket{1}\ket{\phi_1}+c_2\ket{0}\ket{\phi_2}$$
to either $\ket{1}\ket{\phi_1}$ or $\ket{0}\ket{\phi_2}$. Of course, at that point an operator could step up and reprogram the machine, but such an intervention would clearly violate the principle of autonomy of operation of the device in the same way that it would if a desktop computer were unable to perform the NOT operation and had to request the manual flipping of a switch from an operator each time the unary operator were required.

\section{Conclusion}

The {halting scheme for {quantum Turing machine}s}, as proposed by {Deutsch}, is a valid idea---or appears, so far, to be---if one keeps in mind that the operation of the machine will be essentially probabilistic and not deterministic. Any classical reversible universal Turing machine, which uses the halt bit to identify its single terminal state and which after reaching the terminal state keeps moving the head in one direction, corresponds to a well-formed {quantum Turing machine} since its operation consists of a permutation of the basic states of the machine. This provides a simple example of one machine for which the halting scheme is obviously valid. However it is the view of the authors that more research is needed into the power of well-formed {quantum Turing machine}s. Among other things, although a simulation procedure has been described for arbitrary QTMs, it has not been clearly shown whether any universal such machine (or a machine universal for some subclass) exists. The {halting scheme for {quantum Turing machine}s}, while providing the means of using them in practice, seems to be a serious impediment in this regard.

\bibliographystyle{splncs}

\end{document}